\documentclass[12pt]{article}

\begin{document}

\title{{The Construction of Quantum Field Operators: Something of Interest}\thanks{The invited talk at the VIII International Workshop "Applied Category Theory. Graph-Operad-Logic", San Blas, Nayarit, M\'exico, January 9-16, 2010, and at the 6th International Conference on the Dark Side of the Universe (DSU2010), Leon, Gto, M\'exico, June 1-6, 2010.}}

\author{Valeriy V. Dvoeglazov\\
Universidad de Zacatecas\\Ap. Postal 636, Suc. 3 Cruces, C. P. 98062\\Zacatecas, Zac., M\'exico}

\date{June 5, 2010}

\maketitle

\begin{abstract}
We  draw attention to some tune problems in constructions of the quantum-field operators for spins 1/2 and 1. They are related to the existence of negative-energy and acausal solutions of relativistic wave equations. Particular attention is paid to the chiral theories, and to the method of the Lorentz boosts.
\end{abstract}


\newpage

\large{

\section{The Dirac Equation.}

First of all, I would like to remind you some basic things in the quantum field theory.

The Dirac equation has been considered in detail in a pedagogical way~\cite{Sakurai,Ryder}:
\begin{equation}
[i\gamma^\mu \partial_\mu -m]\Psi (x) =0\,.\label{Dirac}
\end{equation}
At least, 3 methods of its derivation exist:
\begin{itemize}
  \item the Dirac one (the Hamiltonian should be linear in $\partial/\partial x^\mu$, and be compatible with $E^2 -{\bf p}^2 c^2 =m^2 c^4$);
  \item the Sakurai one (based on the equation \linebreak $(E- {\bf \sigma} \cdot {\bf p}) (E+ {\bf \sigma} \cdot {\bf p}) \phi =m^2 \phi$);
  \item the Ryder one (the relation between  2-spinors at rest is $\phi_R ({\bf 0}) = \pm \phi_L ({\bf 0})$).
\end{itemize}
The $\gamma^\mu$ are the Clifford algebra matrices 
\begin{equation}
\gamma^\mu \gamma^\nu +\gamma^\nu \gamma^\mu = 2g^{\mu\nu}\,.
\end{equation}
Usually, everybody uses the following definition of the field operator~\cite{Itzykson}:
\begin{equation}
\Psi (x) = \frac{1}{(2\pi)^3}\sum_\sigma \int \frac{d^3 {\bf p}}{2E_p} [ u_\sigma ({\bf p}) a_\sigma ({\bf p}) e^{-ip\cdot x}
+ v_\sigma ({\bf p}) b_\sigma^\dagger ({\bf p})] e^{+ip\cdot x}]\,,
\end{equation}
as given {\it ab initio}.

I studied in the previous 
works~\cite{Dvoeglazov1,Dvoeglazov2,Dvoeglazov3}:
\begin{itemize}
  \item $\sigma \rightarrow h$  (the helicity basis);
  \item  the modified Sakurai derivation (the additional $m_2 \gamma^5$ term in the Dirac equation);
  \item  the derivation of the Barut equation~\cite{Barut} from the first principles, namely 
  based on the generalized Ryder relation, ($\phi_L^h ({\bf 0}) = \hat A \phi_L^{-h\,\ast} ({\bf 0}) + \hat B \phi_L^{h\,\ast} ({\bf 0})$). In fact, we have the second mass state ($\mu$-meson)  from that equation:
  \begin{equation}[i\gamma^\mu \partial_\mu - \alpha \partial_\mu \partial^\mu  /m -\beta] \psi =0\,;\end{equation}  
\item the self/anti-self charge-conjugate Majorana 4-spinors\linebreak 
\cite{Majorana, Bilenky} in the momentum representation.
  \end{itemize}
The Wigner rules~\cite{Wigner} of the Lorentz transformations
for the $(0,S)$ left- $\phi_L ({\bf p})$ and  the $(S,0)$ right-
$\phi_R ({\bf p})$ spinors are:
\begin{eqnarray}
(S,0):&&\phi_R ({\bf p})= \Lambda_R ({\bf p} \leftarrow
{\bf 0})\,\phi_R ({\bf 0})  =  \exp (+\,{\bf S} \cdot
{\bf \varphi}) \,\phi_R ({\bf 0}),\\
(0,S):&&\phi_L ({\bf p}) = \Lambda_L ({\bf p} \leftarrow
{\bf 0})\,\phi_L
({\bf 0})  =  \exp (-\,{\bf S} \cdot {\bf \varphi})\,\phi_L
({\bf 0}),\label{boost0}
\end{eqnarray}
with ${\bf \varphi} = {\bf n} \varphi$ being the boost parameters:
\begin{eqnarray}
&&cosh (\varphi ) =\gamma = \frac{1}{\sqrt{1-v^2/c^2}}, 
sinh (\varphi ) =\beta \gamma =\frac{v/c}{\sqrt{1-v^2/c^2}}\\
&&tanh (\varphi ) =v/c\,.
\end{eqnarray} 
They are well known and given, {\it e.g.}, in~\cite{Wigner,Faustov,Ryder}. 

On using the Wigner rules and the Ryder relations we can recover the Dirac equation in the matrix form:
\begin{eqnarray}
\pmatrix{\mp m \, 1 & p_0 + {\bf \sigma}\cdot {\bf p}\cr
p_0 - {\bf \sigma}\cdot {\bf p} & \mp m \, 1\cr} \psi (p^\mu)\,
=\, 0\,,
\end{eqnarray}
or
$(\gamma\cdot p - m) u ({\bf p})=0$ and $(\gamma\cdot p + m) v ({\bf p})=0$.
We have used the property $\left [\Lambda_{L,R} ({\bf p}
\leftarrow {\bf 0})\right ]^{-1} =
\left [\Lambda_{R,L} ({\bf p} \leftarrow {\bf 0})\right ]^\dagger$ above,
and that both ${\bf S}$ and $\Lambda_{R,L}$ are Hermitian
for the finite $(S=1/2,0)\oplus (0,S=1/2)$ representation
of the Lorentz group.
Introducing $\psi (x) \equiv \psi (p)  \exp (\mp ip\cdot x)$
and letting $p_\mu \rightarrow i\partial_\mu$, the above equation
becomes the Dirac equation (\ref{Dirac}).

The solutions of the Dirac equation are denoted by \linebreak $u ({\bf p}) = column (\phi_R ({\bf p})\quad \phi_L ({\bf p}))$ and $v ({\bf p}) =\gamma^5 u ({\bf p})$.  Let me remind
that the boosted 4-spinors in the common-used basis  (the standard representation of $\gamma$ matrices) are
\begin{eqnarray}
u_{{1\over 2},{1\over 2}} &=& \sqrt{\frac{(E+m)}{2m}}
\pmatrix{1\cr 0\cr p_z/(E+m)\cr p_r/(E+m)\cr}\,,\nonumber\\
u_{{1\over 2},-{1\over 2}} &=&\sqrt{\frac{(E+m)}{2m}}
\pmatrix{0\cr 1\cr p_l/(E+m)\cr -p_z/(E+m)\cr}\,,\\
v_{{1\over 2},{1\over 2}} &=& \sqrt{\frac{(E+m)}{2m}}\pmatrix{p_z/(E+m)\cr p_r/(E+m)\cr
1\cr 0\cr}\,,\nonumber\\
v_{{1\over 2},-{1\over 2}} &=&\sqrt{\frac{(E+m)}{2m}} \pmatrix{p_l/(E+m)\cr -p_z/(E+m)\cr 0\cr 
1\cr}\,.
\end{eqnarray}
$E=\sqrt{{\bf p}^2 +m^2}>0$, $p_0=\pm E$, $p^\pm = E\pm p_z$, $p_{r,l}= p_x\pm ip_y$.
They  are the parity eigenstates with the eigenvalues of $\pm 1$. In
the parity operator the matrix $\gamma_0=\pmatrix{1&0\cr 0 &
-1\cr}$ was used as usual. They also describe
eigenstates of the charge operator, $Q$, if at rest
\begin{equation}\label{rb}
\phi_R ({\bf 0})
=\pm \phi_L ({\bf 0})
\end{equation}
(otherwise the corresponding physical states are no longer
the charge eigenstates). 
Their normalizations are:
\begin{eqnarray}
&&\bar u_\sigma ({\bf p})u_{\sigma^\prime} ({\bf p}) =+\delta_{\sigma\sigma^\prime}\,,\\
&&\bar v_\sigma ({\bf p})v_{\sigma^\prime} ({\bf p}) =-\delta_{\sigma\sigma^\prime}\,,\\
&&\bar u_\sigma ({\bf p})v_{\sigma^\prime} ({\bf p}) = 0\,.
\end{eqnarray}
The bar over the 4-spinors signifies the Dirac conjugation.

Thus in this Section we have used the basis for charged particles in the $(S,0)\oplus (0,S)$ representation (in general) 
\begin{eqnarray}
u_{+\sigma} ({\bf 0}) &=& N(\sigma)\pmatrix{1\cr 0\cr . \cr .
\cr . \cr 0\cr},\,
u_{\sigma-1} ({\bf 0})=N(\sigma) \pmatrix{0\cr
1\cr . \cr .  \cr . \cr 0\cr},\ldots
v_{-\sigma} ({\bf 0})=N(\sigma) \pmatrix{0\cr 0\cr . \cr . \cr . \cr 1\cr}\nonumber\\
&&
\end{eqnarray}
Sometimes, the normalization factor is convenient to choose $N(\sigma)=m^\sigma$ in order
the rest spinors to vanish in the massless limit. 

However, other constructs are possible in the $(1/2,0)\oplus (0,1/2)$ representation.

\section{Majorana Spinors in the Momentum Representation.}

During the 20th century various authors introduced {\it self/anti-self} charge-conjugate 4-spinors
(including in the momentum representation), see~\cite{Majorana,Bilenky,Ziino,Ahluwalia}. 
Later \linebreak\cite{Lounesto,Dvoeglazov1,Dvoeglazov2,Kirchbach} {\it etc} studied these spinors, they found corresponding dynamical equations, gauge transformations 
and other specific features of them.
The definitions are:
\begin{equation}
C= e^{i\theta} \pmatrix{0&0&0&-i\cr
0&0&i&0\cr
0&i&0&0\cr
-i&0&0&0\cr} {\cal K} = -e^{i\theta} \gamma^2 {\cal K}
\end{equation}
is the anti-linear operator of charge conjugation. ${\cal K}$ is the complex conjugation operator. We  define the {\it self/anti-self} charge-conjugate 4-spinors 
in the momentum space
\begin{eqnarray}
C\lambda^{S,A} ({\bf p}) &=& \pm \lambda^{S,A} ({\bf p})\,,\\
C\rho^{S,A} ({\bf p}) &=& \pm \rho^{S,A} ({\bf p})\,.
\end{eqnarray}
Thus,
\begin{equation}
\lambda^{S,A} (p^\mu)=\pmatrix{\pm i\Theta \phi^\ast_L ({\bf p})\cr
\phi_L ({\bf p})}\,,
\end{equation}
and
\begin{equation}
\rho^{S,A} ({\bf p})=\pmatrix{\phi_R ({\bf p})\cr \mp i\Theta \phi^\ast_R ({\bf p})}\,.
\end{equation}
The Wigner matrix is
\begin{equation}
\Theta_{[1/2]}=-i\sigma_2=\pmatrix{0&-1\cr
1&0}\,,
\end{equation}
and $\phi_L$, $\phi_R$ can be boosted with $\Lambda_{L,R}$ 
matrices.\footnote{Such definitions of 4-spinors differ, of course, from the original Majorana definition in x-representation:
\begin{equation}
\nu (x) = \frac{1}{\sqrt{2}} (\Psi_D (x) + \Psi_D^c (x))\,,
\end{equation}
$C \nu (x) = \nu (x)$ that represents the positive real $C-$ parity field operator. However, the momentum-space Majorana-like spinors 
open various possibilities for description of neutral  particles 
(with experimental consequences, see~\cite{Kirchbach}). For instance, "for imaginary $C$ parities, the neutrino mass 
can drop out from the single $\beta $ decay trace and 
reappear in $0\nu \beta\beta $, a curious and in principle  
experimentally testable signature for a  non-trivial impact of 
Majorana framework in experiments with polarized sources."}

The rest $\lambda$ and $\rho$ spinors are:
\begin{eqnarray}
\lambda^S_\uparrow ({\bf 0}) &=& \sqrt{\frac{m}{2}}
\pmatrix{0\cr i \cr 1\cr 0}\,,\,
\lambda^S_\downarrow ({\bf 0})= \sqrt{\frac{m}{2}}
\pmatrix{-i \cr 0\cr 0\cr 1}\,,\,\\
\lambda^A_\uparrow ({\bf 0}) &=& \sqrt{\frac{m}{2}}
\pmatrix{0\cr -i\cr 1\cr 0}\,,\,
\lambda^A_\downarrow ({\bf 0}) = \sqrt{\frac{m}{2}}
\pmatrix{i\cr 0\cr 0\cr 1}\,,\,\\
\rho^S_\uparrow ({\bf 0}) &=& \sqrt{\frac{m}{2}}
\pmatrix{1\cr 0\cr 0\cr -i}\,,\,
\rho^S_\downarrow ({\bf 0}) = \sqrt{\frac{m}{2}}
\pmatrix{0\cr 1\cr i\cr 0}\,,\,\\
\rho^A_\uparrow ({\bf 0}) &=& \sqrt{\frac{m}{2}}
\pmatrix{1\cr 0\cr 0\cr i}\,,\,
\rho^A_\downarrow ({\bf 0}) = \sqrt{\frac{m}{2}}
\pmatrix{0\cr 1\cr -i\cr 0}\,.
\end{eqnarray}
Thus, in this basis the explicite forms of the 4-spinors of the second kind  $\lambda^{S,A}_{\uparrow\downarrow}
({\bf p})$ and $\rho^{S,A}_{\uparrow\downarrow} ({\bf p})$
are
\begin{eqnarray}
\lambda^S_\uparrow ({\bf p}) &=& \frac{1}{2\sqrt{E+m}}
\pmatrix{ip_l\cr i (p^- +m)\cr p^- +m\cr -p_r},
\lambda^S_\downarrow ({\bf p})= \frac{1}{2\sqrt{E+m}}
\pmatrix{-i (p^+ +m)\cr -ip_r\cr -p_l\cr (p^+ +m)}\nonumber\\
\\
\lambda^A_\uparrow ({\bf p}) &=& \frac{1}{2\sqrt{E+m}}
\pmatrix{-ip_l\cr -i(p^- +m)\cr (p^- +m)\cr -p_r},
\lambda^A_\downarrow ({\bf p}) = \frac{1}{2\sqrt{E+m}}
\pmatrix{i(p^+ +m)\cr ip_r\cr -p_l\cr (p^+ +m)}\nonumber\\
\\
\rho^S_\uparrow ({\bf p}) &=& \frac{1}{2\sqrt{E+m}}
\pmatrix{p^+ +m\cr p_r\cr ip_l\cr -i(p^+ +m)},
\rho^S_\downarrow ({\bf p}) = \frac{1}{2\sqrt{E+m}}
\pmatrix{p_l\cr (p^- +m)\cr i(p^- +m)\cr -ip_r}\nonumber\\
\\
\rho^A_\uparrow ({\bf p}) &=& \frac{1}{2\sqrt{E+m}}
\pmatrix{p^+ +m\cr p_r\cr -ip_l\cr i (p^+ +m)},
\rho^A_\downarrow ({\bf p}) = \frac{1}{2\sqrt{E+m}}
\pmatrix{p_l\cr (p^- +m)\cr -i(p^- +m)\cr ip_r}.\nonumber
\\
\end{eqnarray}
As we showed $\lambda$ and $\rho$ 4-spinors are NOT the eigenspinors of the helicity. Moreover, 
$\lambda$ and $\rho$ are NOT the eigenspinors of the parity (in this representation $P=\pmatrix{0&1\cr 1&0}R$), as opposed to the Dirac case.
The indices $\uparrow\downarrow$ should be referred to the chiral helicity 
quantum number introduced 
in the 60s, $\eta=-\gamma^5 h$.
While 
\begin{equation}
Pu_\sigma ({\bf p}) = + u_\sigma ({\bf p})\,,
Pv_\sigma ({\bf p}) = - v_\sigma ({\bf p})\,,
\end{equation}
we have
\begin{equation}
P\lambda^{S,A} ({\bf p}) = \rho^{A,S} ({\bf p})\,,
P \rho^{S,A} ({\bf p}) = \lambda^{A,S} ({\bf p})\,,
\end{equation}
for the Majorana-like momentum-space 4-spinors on the first quantization level.
In this basis one has
\begin{eqnarray}
\rho^S_\uparrow ({\bf p}) \,&=&\, - i \lambda^A_\downarrow ({\bf p})\,,\,
\rho^S_\downarrow ({\bf p}) \,=\, + i \lambda^A_\uparrow ({\bf p})\,,\,\\
\rho^A_\uparrow ({\bf p}) \,&=&\, + i \lambda^S_\downarrow ({\bf p})\,,\,
\rho^A_\downarrow ({\bf p}) \,=\, - i \lambda^S_\uparrow ({\bf p})\,.
\end{eqnarray}

The normalization of the spinors $\lambda^{S,A}_{\uparrow\downarrow}
({\bf p})$ and $\rho^{S,A}_{\uparrow\downarrow} ({\bf p})$ are the following ones:
\begin{eqnarray}
\overline \lambda^S_\uparrow ({\bf p}) \lambda^S_\downarrow ({\bf p}) \,&=&\,
- i m \quad,\quad
\overline \lambda^S_\downarrow ({\bf p}) \lambda^S_\uparrow ({\bf p}) \,= \,
+ i m \quad,\quad\\
\overline \lambda^A_\uparrow ({\bf p}) \lambda^A_\downarrow ({\bf p}) \,&=&\,
+ i m \quad,\quad
\overline \lambda^A_\downarrow ({\bf p}) \lambda^A_\uparrow ({\bf p}) \,=\,
- i m \quad,\quad\\
\overline \rho^S_\uparrow ({\bf p}) \rho^S_\downarrow ({\bf p}) \, &=&  \,
+ i m\quad,\quad
\overline \rho^S_\downarrow ({\bf p}) \rho^S_\uparrow ({\bf p})  \, =  \,
- i m\quad,\quad\\
\overline \rho^A_\uparrow ({\bf p}) \rho^A_\downarrow ({\bf p})  \,&=&\,
- i m\quad,\quad
\overline \rho^A_\downarrow ({\bf p}) \rho^A_\uparrow ({\bf p}) \,=\,
+ i m\quad.
\end{eqnarray}
All other conditions are equal to zero.

The dynamical coordinate-space equations are:
\begin{eqnarray}
i \gamma^\mu \partial_\mu \lambda^S (x) - m \rho^A (x) &=& 0 \,,
\label{11}\\
i \gamma^\mu \partial_\mu \rho^A (x) - m \lambda^S (x) &=& 0 \,,
\label{12}\\
i \gamma^\mu \partial_\mu \lambda^A (x) + m \rho^S (x) &=& 0\,,
\label{13}\\
i \gamma^\mu \partial_\mu \rho^S (x) + m \lambda^A (x) &=& 0\,.
\label{14}
\end{eqnarray}
These are NOT the Dirac equation.
However, they can be written in the 8-component form as follows:
\begin{eqnarray}
\left [i \Gamma^\mu \partial_\mu - m\right ] \Psi_{_{(+)}} (x) &=& 0\,,
\label{psi1}\\
\left [i \Gamma^\mu \partial_\mu + m\right ] \Psi_{_{(-)}} (x) &=& 0\,,
\label{psi2}
\end{eqnarray}
with
\begin{eqnarray}
\Psi_{(+)} (x) = \pmatrix{\rho^A (x)\cr
\lambda^S (x)\cr},
\Psi_{(-)} (x) = \pmatrix{\rho^S (x)\cr
\lambda^A (x)\cr}, \mbox{and}\,\Gamma^\mu =\pmatrix{0 & \gamma^\mu\cr
\gamma^\mu & 0\cr}
\end{eqnarray}
One can also re-write the equations into the two-component form.
Similar formulations have been presented by M. Markov\linebreak \cite{Markov}, and
A. Barut and G. Ziino~\cite{Ziino}. The group-theoretical basis for such doubling has been given
in the papers by Gelfand, Tsetlin and Sokolik~\cite{Gelfand}.

The Lagrangian is
\begin{eqnarray}
&&{\cal L}= \frac{i}{2} \left[\bar \lambda^S \gamma^\mu \partial_\mu \lambda^S - (\partial_\mu \bar \lambda^S ) \gamma^\mu \lambda^S +
\bar \rho^A \gamma^\mu \partial_\mu \rho^A - (\partial_\mu \bar \rho^A ) \gamma^\mu \rho^A +\right.\nonumber\\
&&\left.\bar \lambda^A \gamma^\mu \partial_\mu \lambda^A - (\partial_\mu \bar \lambda^A ) \gamma^\mu \lambda^A +
\bar \rho^S
\gamma^\mu \partial_\mu \rho^S - (\partial_\mu \bar \rho^S ) \gamma^\mu \rho^S -\right.\nonumber\\
&&\left. - m (\bar\lambda^S \rho^A +\bar \lambda^S \rho^A -\bar\lambda^S \rho^A -\bar\lambda^S \rho^A )
\right ]
\end{eqnarray}

The connection with the Dirac spinors has been found. 
For instance,
\begin{eqnarray}
\pmatrix{\lambda^S_\uparrow ({\bf p}) \cr \lambda^S_\downarrow ({\bf p}) \cr
\lambda^A_\uparrow ({\bf p}) \cr \lambda^A_\downarrow ({\bf p})\cr} = {1\over
2} \pmatrix{1 & i & -1 & i\cr -i & 1 & -i & -1\cr 1 & -i & -1 & -i\cr i&
1& i& -1\cr} \pmatrix{u_{+1/2} ({\bf p}) \cr u_{-1/2} ({\bf p}) \cr
v_{+1/2} ({\bf p}) \cr v_{-1/2} ({\bf p})\cr}.\label{connect}
\end{eqnarray}
See also ref.~\cite{Gelfand,Ziino}.

The sets of $\lambda$ spinors and of $\rho$ spinors are claimed to be
{\it bi-orthonormal} sets each in the mathematical sense~\cite{Ahluwalia},  provided
that overall phase factors of 2-spinors $\theta_1 +\theta_2 = 0$ or $\pi$.
For instance, on the classical level $\bar \lambda^S_\uparrow
\lambda^S_\downarrow = 2iN^2 \cos ( \theta_1 + \theta_2 )$.\footnote{We used above 
$\theta_1=\theta_2 =0$.} 

Few remarks have been given in the previous works:
\begin{itemize}
\item
While in the massive case there are four $\lambda$-type spinors, two
$\lambda^S$ and two $\lambda^A$ (the $\rho$ spinors are connected by
certain relations with the $\lambda$ spinors for any spin case),  in a
massless case $\lambda^S_\uparrow$ and $\lambda^A_\uparrow$ identically
vanish, provided that one takes into account that $\phi_L^{\pm 1/2}$ are
 eigenspinors of ${\bf \sigma}\cdot \hat {\bf n}$, the
$2\times 2$ helicity operator.

\item
It was noted the possibility of the generalization of the concept of the
Fock space, which leads to the ``doubling" Fock space~\cite{Gelfand,Ziino}.

\end{itemize}

It was shown~\cite{Dvoeglazov1} that the covariant derivative (and, hence, the
 interaction) can be introduced in this construct in the following way:
\begin{equation}
\partial_\mu \rightarrow \nabla_\mu = \partial_\mu - ig \L^5 A_\mu\quad,
\end{equation}
where $\L^5 = \mbox{diag} (\gamma^5 \quad -\gamma^5)$, the $8\times 8$
matrix. With respect to the transformations

\begin{eqnarray}
\lambda^\prime (x)
\rightarrow (\cos \alpha -i\gamma^5 \sin\alpha) \lambda
(x)\quad,\label{g10}\\
\overline \lambda^{\,\prime} (x) \rightarrow
\overline \lambda (x) (\cos \alpha - i\gamma^5
\sin\alpha)\quad,\label{g20}\\
\rho^\prime (x) \rightarrow  (\cos \alpha +
i\gamma^5 \sin\alpha) \rho (x) \quad,\label{g30}\\
\overline \rho^{\,\prime} (x) \rightarrow  \overline \rho (x)
(\cos \alpha + i\gamma^5 \sin\alpha)\quad\label{g40}
\end{eqnarray}
the spinors retain their properties to be self/anti-self charge conjugate
spinors and the proposed Lagrangian~\cite[p.1472]{Dvoeglazov1} remains to be invariant.
This tells us that while self/anti-self charge conjugate states have
zero eigenvalues of the ordinary (scalar) charge operator but they can
possess the axial charge (cf.  with the discussion of~\cite{Ziino} and
the old idea of R. E. Marshak).

In fact, from this consideration one can recover the Feynman-Gell-Mann
equation (and its charge-conjugate equation). It is re-written in the
two-component form

\begin{eqnarray} 
\cases{\left [\pi_\mu^- \pi^{\mu\,-}
-m^2 -{g\over 2} \sigma^{\mu\nu} F_{\mu\nu} \right ] \chi (x)=0\,, &\cr
\left [\pi_\mu^+ \pi^{\mu\,+} -m^2
+{g\over 2} \widetilde\sigma^{\mu\nu} F_{\mu\nu} \right ] \phi (x)
=0\,, &\cr}\label{iii}
\end{eqnarray}
where already one has $\pi_\mu^\pm =
i\partial_\mu \pm gA_\mu$, \, $\sigma^{0i} = -\widetilde\sigma^{0i} =
i\sigma^i$, $\sigma^{ij} = \widetilde\sigma^{ij} = \epsilon_{ijk}
\sigma^k$ and $\nu^{^{DL}} (x) =\mbox{column} (\chi \quad \phi )$.

Next, because the transformations
\begin{eqnarray}
\lambda_S^\prime ({\bf p}) &=& \pmatrix{\Xi &0\cr 0&\Xi} \lambda_S ({\bf p})
\equiv \lambda_A^\ast ({\bf p}),\\
\lambda_S^{\prime\prime} ({\bf p}) &=& \pmatrix{i\Xi &0\cr 0&-i\Xi} \lambda_S
({\bf p}) \equiv -i\lambda_S^\ast ({\bf p}),\\
\lambda_S^{\prime\prime\prime} ({\bf p}) &=& \pmatrix{0& i\Xi\cr
i\Xi &0\cr} \lambda_S ({\bf p}) \equiv i\gamma^0 \lambda_A^\ast
({\bf p}),\\
\lambda_S^{IV} ({\bf p}) &=& \pmatrix{0& \Xi\cr
-\Xi&0\cr} \lambda_S ({\bf p}) \equiv \gamma^0\lambda_S^\ast
({\bf p})
\end{eqnarray}
with the $2\times 2$ matrix $\Xi$ defined as ($\phi$ is the azimuthal
angle  related with ${\bf p} \rightarrow {\bf 0}$)
\begin{equation}
\Xi = \pmatrix{e^{i\phi} & 0\cr 0 &
e^{-i\phi}\cr}\quad,\quad \Xi \Lambda_{R,L} ({\bf p} \leftarrow
{\bf 0}) \Xi^{-1} = \Lambda_{R,L}^\ast ({\bf p} \leftarrow
 {\bf 0})\,\,\, ,
\end{equation}
and corresponding transformations for
$\lambda^A$ do {\it not} change the properties of bispinors to be in the
self/anti-self charge conjugate spaces, the Majorana-like field operator
($b^\dagger \equiv a^\dagger$) admits additional phase (and, in general,
normalization) transformations:
\begin{equation} \nu^{ML\,\,\prime}
(x^\mu) = \left [ c_0 + i({\bf \tau}\cdot  {\bf c}) \right
]\nu^{ML\,\,\dagger} (x^\mu) \,, 
\end{equation} 
where $c_\alpha$ are
arbitrary parameters. The ${\bf \tau}$ matrices are defined over the
field of $2\times 2$ matrices and the Hermitian
conjugation operation is assumed to act on the $c$- numbers as the complex
conjugation. One can parametrize $c_0 = \cos\phi$ and ${\bf c} = {\bf n}
\sin\phi$ and, thus, define the $SU(2)$ group of phase transformations.
One can select the Lagrangian which is composed from the both field
operators (with $\lambda$ spinors and $\rho$ spinors)
and which remains to be
invariant with respect to this kind of transformations.  The conclusion
is: it is permitted a non-Abelian construct which is based on
the spinors of the Lorentz group only (cf. with the old ideas of T. W.
Kibble and R. Utiyama) .  This is not surprising because both the $SU(2)$
group and $U(1)$ group are  the sub-groups of the extended Poincar\'e group
(cf.~\cite{Ryder}).

The Dirac-like and Majorana-like field operators can
be built from both $\lambda^{S,A} ({\bf p})$ and $\rho^{S,A} ({\bf p})$,
or their combinations. For 
instance,
\begin{eqnarray}
&&\Psi (x^\mu) \equiv \int {d^3 {\bf p}\over (2\pi)^3} {1\over 2E_p}
\sum_\eta \left [ \lambda^S_\eta ({\bf p}) \, a_\eta ({\bf p}) \,\exp
(-ip\cdot x) +\right.\nonumber\\
&+&\left.\lambda^A_\eta ({\bf p})\, b^\dagger_\eta ({\bf p}) \,\exp
(+ip\cdot x)\right ].\label{oper}
\end{eqnarray}

The anticommutation relations are the following ones (due to the {\it bi-orthonormality}):
\begin{eqnarray}
[a_{\eta{\prime}} ({\bf p}^{\prime}), a_\eta^\dagger ({\bf p}) ]_\pm = (2\pi)^3 2E_p \delta ({\bf p} -{\bf p}^\prime) \delta_{\eta,-\eta^\prime}
\end{eqnarray}
and 
\begin{eqnarray}
[b_{\eta{\prime}} ({\bf p}^{\prime}), b_\eta^\dagger ({\bf p}) ]_\pm = (2\pi)^3 2E_p \delta ({\bf p} -{\bf p}^\prime) \delta_{\eta,-\eta^\prime}
\end{eqnarray}
Other (anti)commutators are equal to zero: ($[ a_{\eta^\prime} ({\bf p}^{\prime}), 
b_\eta^\dagger ({\bf p}) ]=0$).

In the Fock space operations of the charge conjugation and space
inversions can be defined through unitary operators such that:
\begin{eqnarray}
U^c_{[1/2]} \Psi (x^\mu) (U^c_{[1/2]})^{-1} &=& {\cal C}_{[1/2]}
\Psi^\dagger_{[1/2]} (x^\mu),\\
U^s_{[1/2]} \Psi (x^\mu) (U^s_{[1/2]})^{-1} &=& \gamma^0
\Psi (x^{\prime^{\,\mu}}),
\end{eqnarray}
the time reversal operation, through {\it an antiunitary}
operator\footnote{Let us remind that the operator of hermitian conjugation does not act on $c$-numbers on the left side of the equation (\ref{tr}).
This fact is conected with the properties of an antiunitary operator:
$\left [ V^{^T} \lambda A (V^{^T})^{-1}\right ]^\dagger =
\left [\lambda^\ast V^{^T} A (V^{^T})^{-1}\right ]^\dagger =
\lambda \left [ V^{^T} A^\dagger (V^{^T})^{-1} \right ]$.}
\begin{equation}
\left [V^{^T}_{[1/2]}  \Psi (x^\mu)
(V^{^T}_{[1/2]})^{-1} \right ]^\dagger = S(T) \Psi^\dagger
(x^{{\prime\prime}^\mu}) \quad,\label{tr}
\end{equation}
with
$x^{\prime^{\,\mu}} \equiv (x^0, -{\bf x})$ and $x^{{\prime\prime}^{\,\mu}}
=(-x^0,{\bf x})$.  We  further assume the vacuum state to be assigned an
even $P$- and $C$-eigenvalue and, then, proceed as in ref.~\cite{Itzykson}.

As a result we have the following properties of creation (annihilation)
operators in the Fock space:
\begin{eqnarray}
U^s_{[1/2]} a_\uparrow ({\bf p}) (U^s_{[1/2]})^{-1} &=& - ia_\downarrow
(-  {\bf p}),\nonumber\\
U^s_{[1/2]} a_\downarrow ({\bf p}) (U^s_{[1/2]})^{-1} &=& + ia_\uparrow
(- {\bf p})\\
U^s_{[1/2]} b_\uparrow^\dagger ({\bf p}) (U^s_{[1/2]})^{-1} &=&
+ i b_\downarrow^\dagger (- {\bf p}),\nonumber\\
U^s_{[1/2]} b_\downarrow^\dagger ({\bf p}) (U^s_{[1/2]})^{-1} &=&
- i b_\uparrow (- {\bf p}),
\end{eqnarray} 
what signifies that the states created by the operators $a^\dagger
({\bf p})$ and $b^\dagger ({\bf p})$ have very different properties
with respect to the space inversion operation, comparing with
Dirac states (the case also regarded in~\cite{Ziino}):
\begin{eqnarray}
U^s_{[1/2]} \vert {\bf p},\,\uparrow >^+ &=& + i \vert -{\bf p},\,
\downarrow >^+ ,
U^s_{[1/2]} \vert {\bf p},\,\uparrow >^- = + i
\vert -{\bf p},\, \downarrow >^-\nonumber\\
\\
U^s_{[1/2]} \vert {\bf p},\,\downarrow >^+ &=& - i \vert -{\bf p},\,
\uparrow >^+,
U^s_{[1/2]} \vert {\bf p},\,\downarrow >^- =  - i
\vert -{\bf p},\, \uparrow >^- .\nonumber\\
\end{eqnarray}

For the charge conjugation operation in the Fock space we have
two physically different possibilities. The first one, {\it e.g.},
\begin{eqnarray}
U^c_{[1/2]} a_\uparrow ({\bf p}) (U^c_{[1/2]})^{-1} &=& + b_\uparrow
({\bf p})\,,\,
U^c_{[1/2]} a_\downarrow ({\bf p}) (U^c_{[1/2]})^{-1} = + b_\downarrow
({\bf p}),\nonumber\\
\\
U^c_{[1/2]} b_\uparrow^\dagger ({\bf p}) (U^c_{[1/2]})^{-1} &=&
-a_\uparrow^\dagger ({\bf p})\,,\,
U^c_{[1/2]} b_\downarrow^\dagger ({\bf p})
(U^c_{[1/2]})^{-1} = -a_\downarrow^\dagger ({\bf p})\,,\nonumber\\
\end{eqnarray}
in fact, has some similarities with the Dirac construct.
However, the action of this operator on the physical states are
\begin{eqnarray}
U^c_{[1/2]} \vert {\bf p}, \uparrow >^+ &=& + \vert {\bf p},
\uparrow >^- ,\,
U^c_{[1/2]} \vert {\bf p}, \downarrow >^+ = + \vert {\bf p},
\downarrow >^- ,\\
U^c_{[1/2]} \vert {\bf p}, \uparrow >^-
&=&  - \, \vert {\bf p}, \uparrow >^+ ,\,
U^c_{[1/2]} \vert
{\bf p}, \, \downarrow >^- = - \vert {\bf p}, \downarrow >^+ .
\end{eqnarray}
But, one can also construct the charge conjugation operator in the
Fock space which acts, {\it e.g.}, in the following manner:
\begin{eqnarray}
\widetilde U^c_{[1/2]} a_\uparrow ({\bf p}) (\widetilde U^c_{[1/2]})^{-1}
&=& - b_\downarrow ({\bf p})\,,\, \widetilde U^c_{[1/2]}
a_\downarrow ({\bf p}) (\widetilde U^c_{[1/2]})^{-1} = - b_\uparrow
({\bf p}),\nonumber\\
\\
\widetilde U^c_{[1/2]} b_\uparrow^\dagger ({\bf p})
(\widetilde U^c_{[1/2]})^{-1} &=& + a_\downarrow^\dagger ({\bf
p})\,,\,
\widetilde U^c_{[1/2]} b_\downarrow^\dagger ({\bf p})
(\widetilde U^c_{[1/2]})^{-1} = + a_\uparrow^\dagger ({\bf p}),\nonumber
\\
\end{eqnarray}
and, therefore,
\begin{eqnarray}
\widetilde U^c_{[1/2]} \vert {\bf p}, \, \uparrow >^+ &=& - \,\vert {\bf
p},\, \downarrow >^- \,,\,
\widetilde U^c_{[1/2]} \vert {\bf p}, \, \downarrow
>^+ = - \, \vert {\bf p},\, \uparrow >^- \,,\\
\widetilde U^c_{[1/2]} \vert
{\bf p}, \, \uparrow >^- &=& + \, \vert {\bf p},\, \downarrow >^+
\,,\,
\widetilde U^c_{[1/2]} \vert {\bf p}, \, \downarrow >^- = + \, \vert {\bf
p},\, \uparrow >^+ \,.
\end{eqnarray}

Investigations of several important cases, which are different from the
 above ones, are required a separate paper to. Next, it is
 possible a situation when the operators of the space inversion and 
charge conjugation commute each other in the Fock space~\cite{Foldy}. For instance,
\begin{eqnarray}
U^c_{[1/2]} U^s_{[1/2]} \vert {\bf
p}, \uparrow >^+ &=& + i U^c_{[1/2]}\vert -{\bf p},\, \downarrow >^+ =
+ i \vert -{\bf p}, \downarrow >^- \\
U^s_{[1/2]} U^c_{[1/2]} \vert {\bf
p}, \uparrow >^+ &=& U^s_{[1/2]}\vert {\bf p}, \uparrow >^- = + i
\vert -{\bf p}, \downarrow >^- .
\end{eqnarray}
The second choice of the charge conjugation operator answers for the case
when the $\widetilde U^c_{[1/2]}$ and $U^s_{[1/2]}$ operations
anticommute:
\begin{eqnarray}
\widetilde U^c_{[1/2]} U^s_{[1/2]} \vert {\bf p},\, \uparrow >^+ &=&
+ i \widetilde U^c_{[1/2]}\vert -{\bf
p},\, \downarrow >^+ = -i \, \vert -{\bf p},\, \uparrow >^- \\
U^s_{[1/2]} \widetilde U^c_{[1/2]} \vert {\bf p},\, \uparrow >^+ &=& -
U^s_{[1/2]}\vert {\bf p},\, \downarrow >^- = + i \, \vert -{\bf p},\,
\uparrow >^- \,.
\end{eqnarray}

Next, one can compose states which would have somewhat similar
properties to those which we have become accustomed.
The states $\vert {\bf p}, \,\uparrow >^+ \pm
i\vert {\bf p},\, \downarrow >^+$ answer for positive (negative) parity,
respectively.  But, what is important, {\it the antiparticle states}
(moving backward in time) have the same properties with respect to the
operation of space inversion as the corresponding {\it particle states}
(as opposed to $j=1/2$ Dirac particles).  
The states which are 
eigenstates of the charge conjugation operator in the Fock space are
\begin{equation}
U^c_{[1/2]} \left ( \vert {\bf p},\, \uparrow >^+ \pm i\,
\vert {\bf p},\, \uparrow >^- \right ) = \mp i\,  \left ( \vert {\bf p},\,
\uparrow >^+ \pm i\, \vert {\bf p},\, \uparrow >^- \right ) \,.
\end{equation}
There is no any simultaneous set of states which would be eigenstates of the 
operator of the space inversion and of the charge conjugation 
$U^c_{[1/2]}$.

Finally, the time reversal {\it anti-unitary} operator in 
the Fock space should be defined in such a way that the formalism to be
 compatible with the $CPT$ theorem. If we wish the Dirac states to transform 
as 
$V(T) \vert {\bf p}, \pm 1/2 > = \pm \,\vert -{\bf p}, \mp 1/2 >$ we
 have to choose (within a phase factor), ref.~\cite{Itzykson}:
\begin{equation}
S(T) = \pmatrix{\Theta_{[1/2]} &0\cr 0 &
\Theta_{[1/2]}\cr}\,.
\end{equation}
Thus, in the first relevant case we obtain for the $\Psi
(x^\mu)$ field, Eq.  (\ref{oper}):
\begin{eqnarray}
V^{^T} a^\dagger_\uparrow ({\bf p}) (V^{^T})^{-1} &=& a^\dagger_\downarrow
(-{\bf p}),\,
V^{^T} a^\dagger_\downarrow ({\bf p}) (V^{^T})^{-1} = -
a^\dagger_\uparrow (-{\bf p}) \\
V^{^T} b_\uparrow ({\bf p}) (V^{^T})^{-1} &=& b_\downarrow
(-{\bf p}),\,
V^{^T} b_\downarrow ({\bf p}) (V^{^T})^{-1} = -
b_\uparrow (-{\bf p}).
\end{eqnarray}
Thus, this construct has very different properties with respect to $C,P$ and $T$ comparing 
with the Dirac construct. 

{\bf But, at least for mathematicians, the dependence of the physical results on the choice 
of the basis is a bit strange thing. Somewhat similar things have been presented in~\cite{Dvoeglazov3} when compared the Dirac-like constructs in the parity and helicity bases. It was shown that the helicity eigenstates
$({\bf \sigma} \cdot {\bf n})\otimes I$) are NOT the parity eigenstates (and the 
${\bf S_3}$ eigenstates), and vice versa, in the helicity basis (cf. with [Berestetskii,Lifshitz, Pitaevskii]), while they obey the same Dirac equation.  The bases are connected by the unitary transformation. And, the both sets of 4-spinors form the complete system in a mathematical sense.}

\section{The Spin 1.}

\subsection{Maxwell Equations as Quantum Equations.}

In refs.~\cite{Gersten,Dvoeglazov4}
the Maxwell-like equations have been derived\footnote{I call them "Maxwell-like"  because an additional gradient of a scalar field $\chi$ can be introduced therein.} from the Klein-Gordon equation. Here they are:
\begin{eqnarray}
&&{\bf \nabla}\times {\bf
E}=-\frac{1}{c}\frac{\partial {\bf B}}{\partial t} + {\bf
\nabla} {\it Im} \chi \,, \label{1A}\\
&&{\bf \nabla }\times {\bf B}=\frac{1}{c}\frac{\partial {\bf
E}}{\partial t}  +{\bf \nabla} {\it Re} \chi\,,\label{2A}\\
&&{\bf \nabla}\cdot {\bf E}=-{1\over c} {\partial \over \partial
t} {\it Re}\chi \,,\label{3}\\
&&{\bf \nabla }\cdot {\bf B}= {1\over
c} {\partial \over \partial t} {\it Im} \chi \,.  \label{4}
\end{eqnarray}
Of course, similar equations can be obtained 
in the massive case $m\neq 0$, i.e., within the Proca-like theory.
We should then consider
\begin{equation}
(E^2 -c^2 {\bf p}^2 - m^2 c^4 ) \Psi^{(3)} =0\, .\label{5}
\end{equation}
In the spin-1/2 case the equation (\ref{5}) can be written 
for the two-component spinor ($c=\hbar =1$)
\begin{equation}
(E I^{(2)} - {\bf\sigma}\cdot {\bf p})
(E I^{(2)} + {\bf\sigma}\cdot {\bf p})\Psi^{(2)} = m^2 \Psi^{(2)}\,,
\end{equation}
or, in the 4-component form
\begin{equation}
[i\gamma_\mu \partial_\mu +m_1 +m_2 \gamma^5 ] \Psi^{(4)} = 0\,.
\end{equation}
In the spin-1 case  we have
\begin{equation}
(E I^{(3)} - {\bf S}\cdot {\bf p})
(E I^{(3)} + {\bf S}\cdot {\bf p}){\bf \Psi}^{(3)} 
- {\bf p} ({\bf p}\cdot {\bf \Psi}^{(3)})= m^2 \Psi^{(3)}\,.
\end{equation}
These lead to (\ref{1A}-\ref{4}), when $m=0$ provided that the $\Psi^{(3)}$ is chosen as a superposition of a vector (the electric field) and an axial vector (the magnetic field).\footnote{We can continue writing down equations for higher spins in a similar fashion.} When $\chi =0$ we recover the common-used Maxwell equations.

Otherwise, we can start with ($c=\hbar=1$)\footnote{The question of both explicite and implicite dependences of the fields on the time
(and, hence, the "whole-partial derivative" )has been studied in~\cite{Brownstein,Dvoeglazov5}.}
\begin{equation}
\frac{\partial {\bf E}}{\partial t} = curl {\bf B}\,,\quad \frac{\partial {\bf B}}{\partial t} = -curl {\bf E}\,.
\end{equation}
Then,
\begin{eqnarray}
\frac{\partial ({\bf E}+i{\bf B})}{\partial t} - curl ({\bf B}-i{\bf E})&=&0\,,\\
\frac{\partial ( {\bf E} -i{\bf B})}{\partial t} -curl ({\bf B} + i{\bf E}) &=&0\,.
\end{eqnarray}
In the component form:
\begin{eqnarray}
\frac{\partial ({\bf E}+i{\bf B})^i}{\partial t} +i\epsilon^{ijk} \partial_j ({\bf E}+i{\bf B})^k &=&0\,,\\
\frac{\partial ( {\bf E} -i{\bf B})^i}{\partial t} - i\epsilon^{ijk}\partial_j ({\bf E} - i{\bf B})^k &=&0\,.
\end{eqnarray}
Since the spin-1 matrices can be presented in the form: $({\bf S}^i)^{jk}= -i \epsilon^{ijk}$,
we  have 
\begin{eqnarray}
\frac{\partial ({\bf E}+i{\bf B})^i}{\partial t} + ({\bf S}\cdot \nabla )^{ik} ({\bf E}+i{\bf B})^k &=&0\,,\\
\frac{\partial ( {\bf E} -i{\bf B})^i}{\partial t} -  ({\bf S}\cdot \nabla )^{ik} ({\bf E} - i{\bf B})^k &=&0\,.
\end{eqnarray}
Finally, on using that $\hat {\bf p} = -i\hbar \nabla$ we have
\begin{equation}
i\frac{\partial \phi}{\partial t} = ({\bf S}\cdot \hat {\bf p}) \phi\,,\quad 
i\frac{\partial \xi}{\partial t} = - ({\bf S}\cdot \hat {\bf p}) \xi\,.
\end{equation} 
In the following we show that these equations can also be considered as the massless limit of the Weinberg $S=1$ 
quantum-field equation.

Meanwhile, we can calculate the determinants  of the above equations, $Det [E \mp ({\bf S}\cdot {\bf p})] =0$, and we can find
that we have both the causal $E=\pm \vert {\bf p}\vert $ and acausal $E=0$ solutions.\footnote{The possible interpretation of the $E=0$ solutions are the stationary fields.} These results  will be useful
in analyzing the spin-1  quantum-field theory below.

\subsection{The Weinberg $2(2S+1)$ Theory for Spin-1}

It is based on the following
postulates~[Wigner,Weinberg]:

\begin{itemize}

\item
The fields transform according to the formula:
\begin{equation}
U [\Lambda, a] \Psi_n (x) U^{-1} [\Lambda, a] = \sum_m D_{nm}
[\Lambda^{-1}] \Psi_m (\Lambda x +a)\,,\label{1} \end{equation}
where $D_{nm} [\Lambda]$ is some representation of $\Lambda$; $x^\mu \rightarrow
\Lambda^\mu_{\quad\nu} \,\,x^\nu +a^\mu$, and $U [\Lambda, a]$ is a
unitary operator.

\item
For $(x-y)$ spacelike one has
\begin{equation}
[\Psi_n (x), \Psi_m (y) ]_\pm =0\,\label{2}
\end{equation}
for fermion and boson fields, respectively.

\item
The interaction Hamiltonian density is said by S. Weinberg to be a scalar,
and it is constructed out of the creation and annihilation operators for
the free particles described by the free Hamiltonian $H_0$.

\item
The $S$-matrix is constructed as an integral of the $T$-ordering
product of the interaction Hamiltonians by the Dyson's formula.

\end{itemize}

In this talk we shall be mainly interested in the free-field theory.
Weinberg wrote: ``In order to discuss theories with parity conservation it
is convenient to use $2(2S+1)$-component fields, like the Dirac field.
These do obey field equations, which can be derived as\ldots consequences
of (\ref{1},\ref{2})."\,\footnote{In the $(2S+1)$ formalism fields
obey only the Klein-Gordon equation, according to the Weinberg wisdom.} 
In such a way he proceeds to form the
$2(2S+1)$-component object
$$\Psi =\pmatrix{\Phi_\sigma\cr \Xi_\sigma\cr}$$
transforming according to the Wigner rules. They are the following ones
(see also above, Eqs. (5,6)):
\begin{eqnarray}
\Phi_\sigma ({\bf p}) &=& \exp (+\Theta \,\hat {\bf p} \cdot {\bf S})
\Phi_\sigma ({\bf 0}) \,,\label{wr1}\\
\Xi_\sigma ({\bf p}) &=& \exp (-\Theta \,\hat {\bf p} \cdot {\bf S})
\Xi_\sigma ({\bf 0}) \,\label{wr2}
\end{eqnarray} 
from the zero-momentum frame. $\Theta$ is the boost parameter,
$\tanh \,\Theta =\vert {\bf p} \vert/ E$, \,$\hat {\bf p} =
{\bf p}/ \vert {\bf p} \vert$, ${\bf p}$ is the 3-momentum of the particle,
${\bf S}$ is the angular momentum operator.
For a given representation the matrices ${\bf S}$ can be constructed. In
the Dirac case (the $(1/2,0)\oplus (0,1/2)$ representation) ${\bf S} =
{\bf \sigma}/2$; in the $S=1$ case (the $(1,0)\oplus (0,1)$
representation) we can choose $(S_i)_{jk} = -i\epsilon_{ijk}$, etc. Hence,
we can explicitly calculate (\ref{wr1},\ref{wr2}).

The task is now to obtain relativistic equations for higher spins.
Weinberg uses the following procedure.
Firstly, he defined the scalar matrix
\begin{equation}
\Pi_{\sigma^\prime \sigma}^{(s)} (q) = (-)^{2s} t_{\sigma^\prime
\sigma}^{\quad \mu_1 \mu_2 \ldots \mu_{2s}} q_{\mu_1} q_{\mu_2}\ldots
q_{\mu_{2s}}
\end{equation}
for the $(S,0)$ representation of the Lorentz group ($q_\mu q_\mu =
-m^2$), with the tensor $t$ being defined by [Weinberg,Eqs.(A4-A5)].
Hence,
\begin{equation}
D^{(s)} [\Lambda] \Pi^{(s)} (q) D^{(s)\,\dagger} [\Lambda] = \Pi^{(s)}
(\Lambda q)\label{wein}
\end{equation}
Since at rest we have $[{\bf S}^{(s)}, \Pi^{(s)} (m)] =0$, then according
to the Schur's lemma $\Pi_{\sigma\sigma^\prime}^{\quad (s)} (m) = m^{2s}
\delta_{\sigma \sigma^\prime}$. After the substitution of $D^{(s)}
[\Lambda]$ in Eq.  (\ref{wein}) one has
\begin{equation}
\Pi^{(s)} (q) = m^{2s} \exp (2\Theta \,\hat {\bf q}
\cdot {\bf S}^{(s)})\,.  \end{equation}
One can construct the analogous
matrix for the $(0,S)$ representation by the same procedure:
\begin{equation} \overline{\Pi}^{(s)} (q) = m^{2s} \exp (- 2\Theta
\hat{\bf q}\cdot {\bf S}^{(s)}) \,.  \end{equation}
Finally, by the direct
verification one has in the coordinate representation
\begin{eqnarray} \overline{\Pi}_{\sigma\sigma^\prime} (-i\partial)
\Phi_{\sigma^\prime} =m^{2s} \Xi_\sigma\,,\\ \Pi_{\sigma\sigma^\prime}
(-i\partial) \Xi_{\sigma^\prime} =m^{2s} \Phi_\sigma\,, \end{eqnarray}
provided that $\Phi_\sigma ({\bf 0})$ and $\Xi_\sigma ({\bf 0})$ are
indistinguishable.\footnote{Later, this fact has been incorporated in the
Ryder book~\cite{Ryder}. Truely speaking, this is an additional postulate.
It is possible that the zero-momentum-frame $2(2S+1)$-component objects
(the 4-spinor in the $(1/2,0)\oplus (0,1/2)$ representation, the bivector
in the $(1,0)\oplus (0,1)$ representation, etc.) are connected by an
arbitrary phase factor~\cite{Dv-ff}.}

As a result one has
\begin{equation}
[ \gamma^{\mu_1 \mu_2 \ldots \mu_{2s}} \partial_{\mu_1} \partial_{\mu_2}
\ldots \partial_{\mu_{2s}} +m^{2s} ] \Psi (x) = 0\,,
\end{equation}
with the Barut-Muzinich-Williams covariantly-defined
matrices \linebreak \cite{Bar-Muz,Sankar}. For the spin-1 they are:
\begin{eqnarray}
&&\gamma_{44} =\pmatrix{0&1\cr 1&0\cr}\,,\quad
\gamma_{i4}=\gamma_{4i} = \pmatrix{0&iS_i\cr
-iS_i & 0\cr}\,,\\
&&\gamma^{ij} = \pmatrix{0&\delta_{ij} -S_i S_j - S_j S_i\cr
\delta_{ij} -S_i S_j - S_j S_i & 0\cr}\, .
\end{eqnarray}
Later Sankaranarayanan and Good considered another version of this 
theory~\cite{Sankar} (see also~\cite{Ahluwalia2}). For the $S=1$ case they introduced the
Weaver-Hammer-Good sign operator, ref.~\cite{Weaver}, $m^{2} \rightarrow
m^{2}\, (i\partial/\partial t)/E$, which led to the different parity
properties of an antiparticle with respect to a {\it boson} particle.
Next,  Tucker and Hammer {\it et al}~\cite{TuckerHammer} introduced another higher-spin
equations. In the spin-1 case it is:
\begin{equation} 
[\gamma_{\mu\nu}
\partial_\mu \partial_\nu + \partial_\mu \partial_\mu -2m^2 ] \Psi^{(s=1)}
= 0\,  
\end{equation}
(Euclidean metric is now used). In fact, they added the Klein-Gordon
equation to the Weinberg equation. One can add the
Klein-Gordon equation with arbitrary multiple factor to the Weinberg
equation. So, we can study the generalized Weinberg-Tucker-Hammer equation
($S=1)$, which is written ($p_\mu = -i\partial/\partial x^\mu$):
\begin{equation}
[\gamma_{\alpha\beta}p_\alpha p_\beta +A p_\alpha p_\alpha +Bm^2 ]
\Psi =0\,.
\end{equation}
It has solutions with relativistic dispersion relations $E^2 -{\bf
p}^2 = m^2$, ($c=\hbar=1$) provided that
\begin{equation}
{B\over A+1} = 1\,, \qquad \mbox{or} \qquad{B\over A-1} =1\,.\label{se}
\end{equation}
This can be proven by considering the algebraic equation \linebreak
$Det [\gamma_{\alpha\beta} p_\alpha p_\beta +A p_\alpha p_\alpha +Bm^2 ]
=0$. It is  of the 12th order in $p_\mu$. Solving it with respect to
energy one obtains the conditions (\ref{se}). {\bf Unlike the Maxwell equations there are
NO any $E=0$ solutions.}

The solutions in the momentum representation have been explicitly presented by~\cite{Ahluwalia2}:
\begin{eqnarray}
u_{+1} ({\bf p})&=&
\pmatrix{m+\left [ (2p_z^2+p_{+} p_{-}) / 2(E+m)\right ]\cr
                      {p_z p_{+}/{\sqrt 2}(E+m)}\cr
              { p_{+}^2/ 2(E+m) }\cr
               p_z\cr
                   {p_{+}/{\sqrt 2}}\cr
                   0\cr}\,,\\
u_{0}({\bf p}) &=& \pmatrix{{p_z p_{-}/{\sqrt 2}(E+m)}\cr
                      m+\left [ {p_{+} p_{-}/(E+m) }\right ]\cr
                       -{p_z p_{+}/{\sqrt 2}(E+m)}\cr
                       {p_{-}/{\sqrt 2}}\cr
                          0\cr
                        {p_{+}/{\sqrt 2}}\cr}\,,\\
u_{-1}({\bf p})&=&\pmatrix{ { p_{-}^2/ 2(E+m) }\cr
                             -{p_z p_{-}/{\sqrt 2}(E+m)}\cr
                   m+\left [ {(2p_z^2+p_{+} p_{-})/ 2(E+m)}\right]\cr
                      0\cr
                      {p_{-}/{\sqrt 2}}\cr
                   -p_z\cr}\,,\label{uv}
\end{eqnarray}
and
\begin{equation}
v_\sigma ({\bf p}) =\gamma_5 u_\sigma ({\bf p}) =\pmatrix{0&1\cr
1&0\cr} U_\sigma ({\bf p})
\end{equation}
in the standard representation of $\gamma_{\mu\nu}$ matrices.
If the 6-component $v ({\bf p})$ are defined in such way, we inevitably would get 
the additional energy-sign operator~\cite{Weaver,Sankar} $\epsilon = i\partial_t /E =\pm 1$ 
in the dynamical equation,
and the different parities of the corresponding boson and antiboson, $\hat P u_\sigma ({\bf p})
= + u_\sigma ({\bf p})$ and $\hat P v_\sigma ({\bf p}) = - v_\sigma ({\bf p})$.

\section{The Construction of Field Operators.}

The method for constructions of field operators has been given in~\cite{Bogoliubov}:\footnote{In this book a bit different  notation for positive- (negative-) energy solutions has been used
comparing with the general accepted one.}
\begin{equation}
\phi (x) = \frac{1}{(2\pi)^{3/2}}\int dk  e^{ikx}\tilde \phi (k)\,.
\end{equation}
From the Klein-Gordon equation we know:
\begin{equation}
(k^2 -m^2) \tilde \phi (k) =0\,.\label{KG1}
\end{equation}
Thus,
\begin{equation}
\tilde \phi (k) = \delta (k^2 - m^2) \phi (k)\,.
\end{equation}
Next,
\begin{eqnarray}
&&\phi (x) = \frac{1}{(2\pi)^{3/2}} \int d k \, e^{ikx}\delta (k^2 -m^2) (\theta (k_0)+ 
\theta (-k_0)) 
\phi (k) =\nonumber\\
&=& \frac{1}{(2\pi)^{3/2}} \int d k \left [ e^{ikx} \delta (k^2 -m^2) \phi^+ (k)
+ e^{-ikx} \delta (k^2 -m^2) \phi^- (k)\right ],\nonumber\\
\end{eqnarray}
where
\begin{equation}
\phi^+ (k) =\theta (k_0) \phi (k)\,,\mbox{and}\,\, \phi^- (k)=\theta (k_0) \phi (-k)\,.
\end{equation}
\begin{eqnarray}
\phi^+ (x) &=& \frac{1}{(2\pi)^{3/2}}\int \frac{d^3 {\bf k}}{2E_k}  e^{+ikx} \phi^+ (k)\,,\\
\phi^- (x) &=& \frac{1}{(2\pi)^{3/2}}\int \frac{d^3 {\bf k}}{2E_k}  e^{-ikx} \phi^- (k)\,.
\end{eqnarray}

In the spinor case (the $(1/2,0)\oplus (0,1/2)$ representation space) we have more components.
Instead of the equation (\ref{KG1}) we have 
\begin{equation}
(\hat k + m) \psi (k) \vert_{k^2 =m^2} =0\,.
\end{equation}
However, again
\begin{equation}
\psi (x) = \frac{1}{(2\pi)^{3/2}} \int d k \, e^{ikx}\delta (k^2 -m^2) (\theta (k_0)+ 
\theta (-k_0)) 
\psi (k)\,,
\end{equation}
and
\begin{equation}
\psi (x)={1\over (2\pi)^3} \int \frac{d^3 {\bf k}}{2E_k} \left [ e^{ikx}\theta(k_0)  
\psi (k)   + e^{-ikx} \theta (k_0) \psi (-k) \right ]\,,
\end{equation}
where $k_0 =E =\sqrt{{\bf k}^2 +m^2}$ is positive in this case. Hence:
\begin{equation}
(\hat k +m)\psi^+ ({\bf k}) =0\,,\quad
(-\hat k + m)\psi^- ({\bf k}) =0\,.
\end{equation}

{\bf Everything is OK? However, please note that the momentum-space Dirac equations
$(\hat k - m) u =0$, $(\hat k +m) v=0$
have solutions $k_0=\pm \sqrt{{\bf k}^2 +m^2}$, both for $u-$ and $v-$ spinors.
This can be checked by calculating the determinants. Usually, one chooses 
$k_0=E=\sqrt{{\bf k}^2 +m^2}$ in the $u-$ and in the $v-$. This is because 
on the classical level (better to say, on the first quantization level) 
the negative-energy $u-$ can be transformed in the positive-energy $v-$, 
and vice versa. This is not precisely so, if we go to the secondary quantization level.
The introduction of creation/annihilation noncommutating operators  gives us more possibilities
in constructing generalized theory even on the basis of the Dirac equation.}

Various-type field operators are possible in the $(1/2,1/2)$ representation. 
During the calculations below we have to present $1=\theta (k_0) +\theta (-k_0)$
( as previously ) in order to get positive- and negative-frequency parts. 
\begin{eqnarray}
&&A_\mu (x) = {1\over (2\pi)^3} \int d^4 k \,\delta (k^2 -m^2) e^{+ik\cdot x}
A_\mu (k) =\nonumber\\
&=& {1\over (2\pi)^3} \sum_{\lambda}^{}\int d^4 k \delta (k_0^2 -E_k^2) e^{+ik\cdot x}
\epsilon_\mu (k,\lambda) a_\lambda (k) =\nonumber\\
&=&{1\over (2\pi)^3} \int {d^4 k \over 2E} [\delta (k_0 -E_k) +\delta (k_0 +E_k) ] 
[\theta (k_0) +\theta (-k_0) ]\nonumber\\
&&e^{+ik\cdot x}
A_\mu (k) = {1\over (2\pi)^3} \int {d^4 k \over 2E} [\delta (k_0 -E_k) +\delta (k_0 +E_k) ] \nonumber\\
&&\left
[\theta (k_0) A_\mu (k) e^{+ik\cdot x}  + 
\theta (k_0) A_\mu (-k) e^{-ik\cdot x} \right ]  =\\
&=&{1\over (2\pi)^3} \int {d^3 {\bf k} \over 2E_k} \theta(k_0)  
[A_\mu (k) e^{+ik\cdot x}  + A_\mu (-k) e^{-ik\cdot x} ]
=\nonumber\\
&=&{1\over (2\pi)^3} \sum_{\lambda}^{}\int {d^3 {\bf k} \over 2E_k}   
[\epsilon_\mu (k,\lambda) a_\lambda (k) e^{+ik\cdot x}  + \epsilon_\mu (-k,\lambda) 
a_\lambda (-k) e^{-ik\cdot x} ].\nonumber
\end{eqnarray}

{\bf In general, due to theorems for integrals and for distributions the presentation $1=\theta (k_0)
+ \theta (-k_0)$ is possible because we use this in the integrand. However, remember, that
we have the $k_0=E=0$ solution of the Maxwell equations.\footnote{Of course, the same procedure can be applied in the construction of the quantum field operator for $F_{\mu\nu}$.} Moreover, it has the experimental confirmation (for instance, the stationary magnetic field $curl {\bf B}=0$). Meanwhile the $theta$ function is NOT defined in $k_0=0$. Do we not loose this solution in the above construction
of the quantum field operator? Mathematicians did not answer me in a straightforward way.}

Moreover, we should transform the second part to $\epsilon_\mu^\ast (k,\lambda) b_\lambda^\dagger (k)$ as usual. In such a way we obtain the charge-conjugate states.\footnote{In the cirtain basis 
it is considered that the charge conjugation operator is just the complex conjugation operator for 4-vectors $A_\mu$.} Of course, one can try to get $P$-conjugates or $CP$-conjugate states too. 

In the Dirac case we should assume the following relation in the field operator:
\begin{equation}
\sum_{\lambda}^{} v_\lambda (k) b_\lambda^\dagger (k) = \sum_{\lambda}^{} u_\lambda (-k) a_\lambda (-k)\,.\label{dcop}
\end{equation}
We know that~\cite{Ryder,Itzykson}
\begin{eqnarray}
\bar u_\mu (k) u_\lambda (k) &=& +m \delta_{\mu\lambda}\,,\\
\bar u_\mu (k) u_\lambda (-k) &=& 0\,,\\
\bar v_\mu (k) v_\lambda (k) &=& -m \delta_{\mu\lambda}\,,\\
\bar v_\mu (k) u_\lambda (k) &=& 0\,,
\end{eqnarray}
but we need $\Lambda_{\mu\lambda} (k) = \bar v_\mu (k) u_\lambda (-k)$.
By direct calculations,  we find
\begin{equation}
-mb_\mu^\dagger (k) = \sum_{\nu}^{} \Lambda_{\mu\lambda} (k) a_\lambda (-k)\,.
\end{equation}
Hence, $\Lambda_{\mu\lambda} = -im ({\bf \sigma}\cdot {\bf n})_{\mu\lambda}$
and 
\begin{equation}
b_\mu^\dagger (k) = i({\bf\sigma}\cdot {\bf n})_{\mu\lambda} a_\lambda (-k)\,.
\end{equation}
Multiplying (\ref{dcop}) by $\bar u_\mu (-k)$ we obtain
\begin{equation}
a_\mu (-k) = -i ({\bf \sigma} \cdot {\bf n})_{\mu\lambda} b_\lambda^\dagger (k)\,.
\end{equation}
Thus, the  above equations  are self-consistent.

In the $(1,0)\oplus (0,1)$ representation we have somewhat different situation. Namely,
\begin{equation}
a_\mu (k) = [1-2({\bf S}\cdot {\bf n})^2]_{\mu\lambda} a_\lambda (-k)\,. 
\end{equation}
This signifies that in order to construct the Sankaranarayanan-Good field operator (which was used by Ahluwalia, Johnson and Goldman~\cite{Ahluwalia2}, it satisfies 
$[\gamma_{\mu\nu} \partial_\mu \partial_\nu - {(i\partial/\partial t)\over E} 
m^2 ] \Psi =0$, we need additional postulates.

We can set for the 4-vector field operator:
\begin{equation}
\sum_{\lambda}^{} \epsilon_\mu (-k,\lambda) a_\lambda (-k) = 
\sum_{\lambda}^{} \epsilon_\mu^\ast (k,\lambda) b_\lambda^\dagger (k)\,,
\label{expan}
\end{equation}
multiply both parts by $\epsilon_\nu [\gamma_{44}]_{\nu\mu}$, and use the normalization conditions for polarization vectors.

However, in the $({1\over 2}, {1\over 2})$ representation we can also expand
(apart the equation (\ref{expan})) in the different way:
\begin{equation}
\sum_{\lambda}^{} \epsilon_\mu (-k, \lambda) a_\lambda (-k) =
\sum_{\lambda}^{} \epsilon_\mu (k, \lambda) a_\lambda (k)\,.
\end{equation}
From the first definition we obtain (the signs $\mp$
depends on the value of $\sigma$):
\begin{equation}
b_\sigma^\dagger (k) = \mp \sum_{\mu\nu\lambda}^{} \epsilon_\nu (k,\sigma) 
[\gamma_{44}]_{\nu\mu} \epsilon_\mu (-k,\lambda) a_\lambda (-k)\,,
\end{equation}
or
\begin{eqnarray}
&&b_\sigma^\dagger (k) = \\
&&{E_k^2 \over m^2} \pmatrix{1+{{\bf k}^2\over E_k^2}&\sqrt{2}
{k_r \over E_k}&-\sqrt{2} {k_l \over E_k}& -{2k_3 \over E_k}\cr
-\sqrt{2} {k_r \over E_k}&-{k_r^2 \over {\bf k}^2}& -{m^2k_3^2\over E_k^2 {\bf k}^2}
+{k_r k_l \over E_k^2} & {\sqrt{2} k_3 k_r \over {\bf k}^2}\cr
\sqrt{2} {k_l \over E_k}&-{m^2 k_3^2 \over E_k^2 {\bf k}^2} + {k_r k_l \over E_k^2}& -{k_l^2\over {\bf k}^2} & -{\sqrt{2} k_3 k_l \over {\bf k}^2}\cr
{2k_3 \over E_k}&{\sqrt{2}k_3 k_r \over {\bf k}^2}& -{\sqrt{2} k_3 k_l\over {\bf k}^2} & {m^2 \over E_k^2} -{2 k_3 \over {\bf k}^2}\cr}
\pmatrix{a_{00} (-k)\cr a_{11} (-k)\cr
a_{1-1} (-k)\cr a_{10} (-k)\cr}\,.\nonumber
\end{eqnarray}

From the second definition $\Lambda^2_{\sigma\lambda} = \mp \sum_{\nu\mu}^{} \epsilon^{\ast}_\nu (k, \sigma) [\gamma_{44}]_{\nu\mu}
\epsilon_\mu (-k, \lambda)$ we have:
\begin{eqnarray}
a_\sigma (k) =  \pmatrix{-1&0&0&0\cr
0&{k_3^2 \over {\bf k}^2}& {k_l^2\over {\bf k}^2} & {\sqrt{2} k_3 k_l \over {\bf k}^2}\cr
0&{k_r^2 \over {\bf k}^2}& {k_3^2\over {\bf k}^2} & -{\sqrt{2} k_3 k_r \over {\bf k}^2}\cr
0&{\sqrt{2}k_3 k_r \over {\bf k}^2}& -{\sqrt{2} k_3 k_l\over {\bf k}^2} & 1-{2 k_3^2 \over {\bf k}^2}\cr}\pmatrix{a_{00} (-k)\cr a_{11} (-k)\cr
a_{1-1} (-k)\cr a_{10} (-k)\cr}.
\end{eqnarray}
It is the strange case: the field operator will only destroy particles (like in the $(1,0)\oplus (0,1)$ case). Possibly, we should think about modifications of the Fock space in this case, or introduce several field operators for the $({1\over 2}, {1\over 2})$ representation.

{\bf However,  other way is possible: to construct the left- and right- parts of the $(1,0)\oplus (0,1)$ field operator 
separately each other. In this case the commutation relations may be more complicated.}

Finally, going back to the rest $(S,0)\oplus (0,S)$ objects. {\bf Bogoliubov
constructs them introducing the products with $delta$ functions like $\delta (k_0 -m)$. 
Then, he makes the boost of the "spinors" only, and changes by hand the $\delta$ to
$\delta (k^2 -m^2)$ (where we already have $k_0 =E=\sqrt{{\bf k}^2 +m^2}$). Mathematicians did not answer me, how can it be possible to make the boost
of the $delta$ functions consistently in such a way.}


\medskip

The conclusion is: we still have few questions unsolved in the bases of the quantum field theory, which open a room for generalized theories.

}

\end{document}